\documentclass[conference,a4paper]{IEEEtran}
\IEEEoverridecommandlockouts
\usepackage[utf8]{inputenc}
\usepackage{amsmath}
\usepackage{amsfonts}
\usepackage{amssymb}
\usepackage[english]{babel}
\usepackage{graphicx}
\usepackage{comment}
\usepackage{cite}
\usepackage{cmap}
\usepackage{setspace}
\usepackage{textgreek} 
\usepackage{tabularx,ragged2e,booktabs}
\usepackage{caption}
\usepackage{subcaption}
\usepackage{tikz}
\usepackage{footnote}
\usepackage{multirow}
\usepackage{url}

\usetikzlibrary{shapes.multipart}
\makesavenoteenv{tabular}
\newcolumntype{L}{>{\RaggedRight\arraybackslash}X} 

\def\BibTeX{{\rm B\kern-.05em{\sc i\kern-.025em b}\kern-.08em T\kern-.1667em\lower.7ex\hbox{E}\kern-.125emX}}

\usepackage{textcomp}
\usepackage{lipsum}

\usepackage{bm}

\newcommand\copyrighttext{%
	\footnotesize \textcopyright \enspace 2022 IEEE. Personal use of this material is permitted. Permission from IEEE must be obtained for all other uses, in any current or future media, including reprinting/republishing this material for advertising or promotional purposes, creating new collective works, for resale or redistribution to servers or lists, or reuse of any copyrighted component of this work in other works.
	}
\newcommand\copyrightnotice{%
\begin{tikzpicture}[remember picture,overlay]
	\node[anchor=south] at (current page.south) {\fbox{\parbox{\dimexpr\textwidth-\fboxsep-\fboxrule\relax}{\copyrighttext}}};
\end{tikzpicture}%
}

\begin{document}

\title{Providing High Capacity for AR/VR traffic \\ in 5G Systems with Multi-Connectivity} 

\author{\IEEEauthorblockN{Maxim Susloparov\IEEEauthorrefmark{1}\IEEEauthorrefmark{2}, Artem Krasilov\IEEEauthorrefmark{1}, Evgeny Khorov\IEEEauthorrefmark{1} } 
	
	\IEEEauthorblockN{\IEEEauthorrefmark{1}Institute for Information Transmission Problems, Russian Academy of Sciences, Moscow, Russia\\
		\IEEEauthorrefmark{2}National Research University Higher School of Economics, Moscow, Russia}
	
	\IEEEauthorblockN{Email:  \{susloparov, krasilov, khorov\}@wireless.iitp.ru}
}
\maketitle

\copyrightnotice

\begin{abstract}
	Augmented and Virtual Reality (AR/VR) is often called a ``killer'' application of 5G systems because it imposes very strict Quality of Service (QoS) requirements related to throughput, latency, and reliability. A high-resolution AR/VR flow requires a bandwidth of dozens of MHz. Since the existing low-frequency bands (i.e., below 6 GHz) have limited bandwidth and are overpopulated, one of the ways to satisfy high AR/VR demands is to use wide frequency channels available in the millimeter-Wave (mmWave) band. However, transmission in the mmWave band suffers from high throughput fluctuation and even blockage, which leads to violation of strict AR/VR latency and reliability requirements. To address this problem, 5G specifications introduce a Multi-Connectivity (MC) feature that allows a mobile user to connect simultaneously to several base stations. The paper considers a scenario with two base stations: the first base station operates in the low-frequency band to provide reliable data delivery, while the second one operates in the mmWave band and offers high data rates when the channel conditions are favorable. An open question that falls out of the scope of specifications is how to balance AR/VR traffic between two links with different characteristics. The paper proposes a Delay-Based Traffic Balancing (DBTB) algorithm that minimizes resource consumption of the low-frequency link while satisfying strict AR/VR QoS requirements. With extensive simulations, DBTB is shown to double the network capacity for AR/VR traffic compared with the state-of-the-art traffic balancing algorithms.         
\end{abstract}


\section{Introduction}
\label{sec:intro}

Augmented and Virtual Reality (AR/VR) are rapidly evolving technologies that established a market of \$$209$B in $2022$~\cite{vr-market-share}. AR/VR is used in various areas, such as gaming and entertainment, education, medicine, driving and aviation training, etc. AR/VR applications generate a high-resolution video stream that is played back on an AR/VR headset. To reduce the complexity and the power consumption of AR/VR headsets, a Cloud AR/VR technology has been proposed~\cite{huawei-cloud-vr}. With this technology, an AR/VR video stream is generated and encoded with popular video codecs (e.g., H.265~\cite{itu-h265}) on a high-performance remote server located on the Internet. The video stream is then transmitted to an AR/VR headset that renders and plays back the obtained video\cite{vr21}. 

Typically, an AR/VR headset is connected to a network with a wired or a Wi-Fi connection~\cite{huawei-cloud-vr}. However, these options greatly limit user mobility as they offer local area coverage (e.g., a room in an apartment). One of the ways to provide more ``freedom'' to the AR/VR user is to use cellular systems.
Note that AR/VR traffic imposes very strict  Quality of Service (QoS) requirements regarding throughput, latency, and frame loss rate. For example, as shown in~\cite{huawei-cloud-vr},  a 4K AR/VR stream requires a throughput above $60$ Mbps, a packet delivery time below $15$ ms, and a video frame loss ratio below  $1$\%. Existing 4G systems cannot satisfy such requirements because their very rigid design \cite{arbat} results in  significant data transmission delays exceeding $20$~ms in commercial networks~\cite{cisco-traffic}. 

5G systems can provide much higher capacity by using wide channels available at high frequencies. Specifically, the 5G New Radio (NR) technology is designed to operate at two Frequency Ranges (FRs): (i) FR1 contains frequencies below $6$~GHz, which are traditionally used in cellular systems, (ii) FR2 contains frequencies from $24$ to $53$~GHz and is called the mmWave range. The base station (gNB) operating in FR2 can use channels as wide as $1$~GHz and, thus, provide multi-Gbps data rates. Several papers study the feasibility of using mmWave links for AR/VR traffic delivery~\cite{reliability-vr-hetnets,multiconnect-nr-dc,cellular-wireless-vr}. 
They show that the mmWave link suffers from high throughput fluctuation and link blockage, which leads to a violation of strict AR/VR latency and reliability requirements.

\begin{figure}[t]
	\centering
	\includegraphics[width=0.45\linewidth]{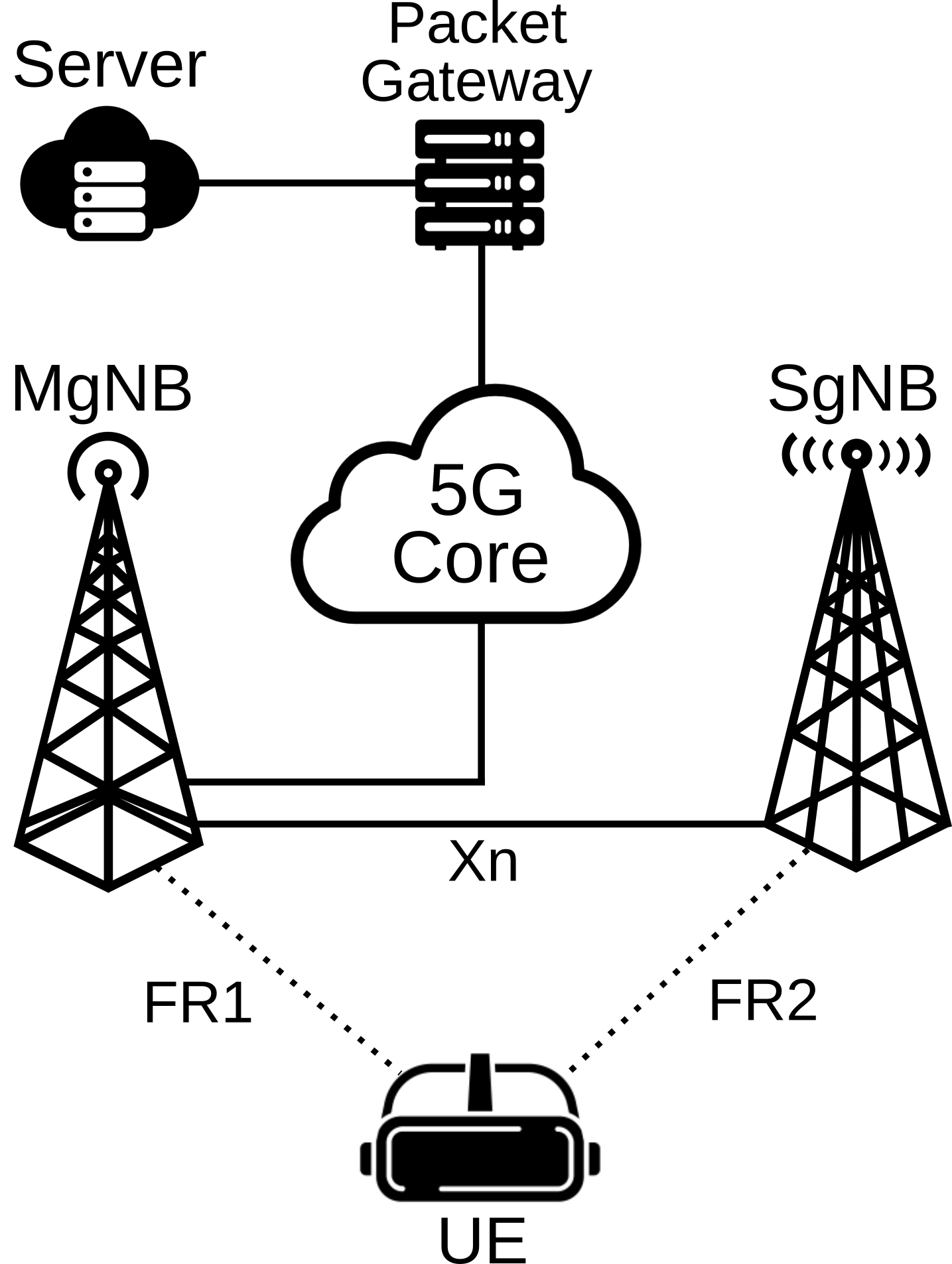} 
	\caption{5G cellular system with the MC feature: MgNB operates in FR1, SgNB operates in FR2.}
	\label{fig:dc-scheme}
\end{figure}

To improve reliability, the NR specifications of Release 16 introduce the Multi-Connectivity (MC) feature, which allows a User Equipment (UE) to connect to several gNBs. 
In this paper, we consider the case when a UE connects to two gNBs called Master gNB (MgNB) and Secondary gNB (SgNB), see Fig.~\ref{fig:dc-scheme}. The MgNB operates in FR1 and, thus,  provides high coverage and reliability. The SgNB operates in FR2 and provides high data rates when channel conditions are favorable. When channel conditions deteriorate (e.g., because of a high probability of mmWave signal blockage), the data rate can be even lower than in FR1~\cite{mmwavek} and no link adaptation algorithms such as \cite{conserv} can improve the situation. In such a case, the UE should use the FR1 link for data transmission. With MC, downlink traffic arrives at the MgNB, which serves as a coordinator. For each packet, the MgNB decides whether to send it via the FR1 link, via the FR2 link, or via both links for better reliability. Thus, an open research question left out of the scope of 5G specifications is how to balance the traffic to utilize several wireless links efficiently.

Most of the existing papers consider the problem of balancing enhanced Mobile Broadband (eMBB) traffic. The algorithms proposed in these papers aim at maximizing the total throughput~\cite{improved-handover-dc-5g-mmwave, dc-handover-ultradense-mmwave, handover-mgmt-camera-images, multi-connect-mobile-networks, energy-efficient-mc}. In this paper, we show that these algorithms cannot be applied to AR/VR traffic because they do not take into account strict latency and reliability requirements. Consequently, we propose a Delay-Based Traffic Balancing (DBTB) algorithm that minimizes resource consumption of the low-frequency link (i.e., the FR1 link) while satisfying strict AR/VR QoS requirements. With extensive simulations in NS-3, we show that DBTB doubles the network capacity for AR/VR traffic compared with the state-of-the-art traffic balancing algorithms.         


The rest of the paper is organized as follows. In Section~\ref{sec:existing-solutions}, we review the existing traffic balancing algorithms designed for 5G systems with the MC feature. Section \ref{sec:problem} provides the problem statement. In Section~\ref{sec:proposed-solution}, we describe the proposed algorithm. In Section~\ref{sec:performance-evaluation}, we compare the performance of the proposed and the existing algorithms in scenarios with AR/VR traffic. Section~\ref{sec:outro} concludes the paper.

\section{Existing algorithms}
\label{sec:existing-solutions}

\begin{figure}[t]
	\centering
	\includegraphics[width=0.5\linewidth]{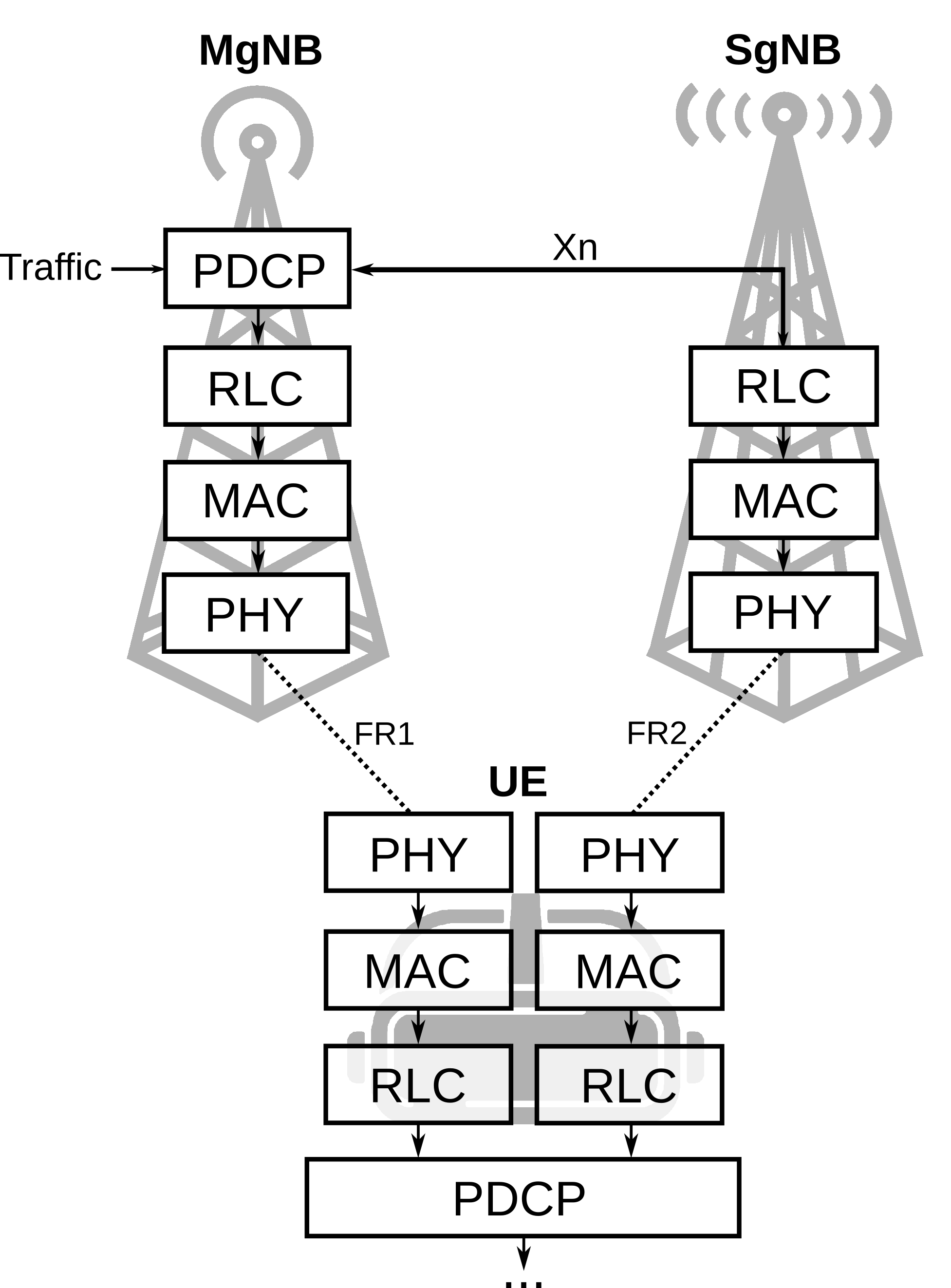} 
	\caption{NR protocol stack for a UE connected to two gNBs.}
	\label{fig:protocol-stack}
\end{figure}

Consider a scenario with a UE connected to two gNBs: an MgNB operating in FR1, and an SgNB operating in FR2. Figure~\ref{fig:protocol-stack} shows the protocol stack of the gNBs and the UE. With MC, the downlink traffic arrives at the Packet Data Convergence Protocol (PDCP) located at the MgNB. With each data packet called Service Data Unit (SDU), PDCP makes the following changes: it (i)~compresses upper-layer (e.g., IP/UDP) headers,  (ii)~chippers data, (iii)~adds a PDCP header which contains a unique sequence number. An SDU with the appended header is called Protocol Data Unit (PDU). With each PDCP PDU, the MgNB determines whether to send it (i)~via the local protocol stack (i.e., via the FR1 link), (ii)~via the SgNB stack (via the FR2 link) by using the Xn interface, or (iii)~via both links. The latter case is called Packet Duplication (PD). When it is used, several copies of the same PDCP PDU may arrive at the UE. To remove duplicates and provide in-sequence packet delivery, the UE uses sequence numbers in the PDCP header~\cite{ts38.323}.

The algorithm used by PDCP to balance the traffic between various links is not specified in 5G standards and shall be defined by vendors. Below we describe the existing algorithms. 



\subsection{Link Switching}

With Link Switching, the MgNB uses one link at a time. To select a link for data transmission, the MgNB uses one or several metrics to choose the best link. 
The widely used metrics are as follows.
\begin{itemize}
	\item Reference Signal Received Power (RSRP) or Signal Interference plus Noise Ratio (SINR)~\cite{improved-handover-dc-5g-mmwave,dc-handover-ultradense-mmwave}. These metrics are updated at the MgNB and the SgNB based on periodic channel measurements.
	\item Expected data rate~\cite{energy-efficient-mc} that can be estimated, e.g., based on the Shannon equation: $C = B \log_2{(1 + SINR)}$, where $B$ is the bandwidth of the considered link. 
\end{itemize}

If the link quality significantly changes with time, the packets already being transmitted via the link may be lost, which is crucial for AR/VR applications. 

\subsection{Packet Duplication}

With Packet Duplication (PD), all data packets are duplicated at PDCP and sent via both links. This approach allows achieving high reliability~\cite{packet-duplication, dc-handover-ultradense-mmwave, cellular-wireless-vr}. However, it significantly increases channel resource consumption, which is critical for the FR1 link because it has much lower bandwidth than the FR2 link.

\subsection{Packet Splitting}

With Packet Splitting (PS), both links are used but packets are distributed between the links without duplication. 
In this paper, we consider a variant of the PS algorithm which provides the fractions of packets sent via each link proportionally to the capacities of the links~\cite{dc-data-split}. Assuming that the link capacities do not vary significantly with time, this approach allows delivering a chunk of data (e.g., a video frame) with minimal time. However, as we show in Section~\ref{sec:performance-evaluation}, this assumption does not hold for FR2 links, which leads to violation of the video frame delay budget in some cases.  

\begin{figure*}[t]
	\centering
	\includegraphics[width=0.7\linewidth]{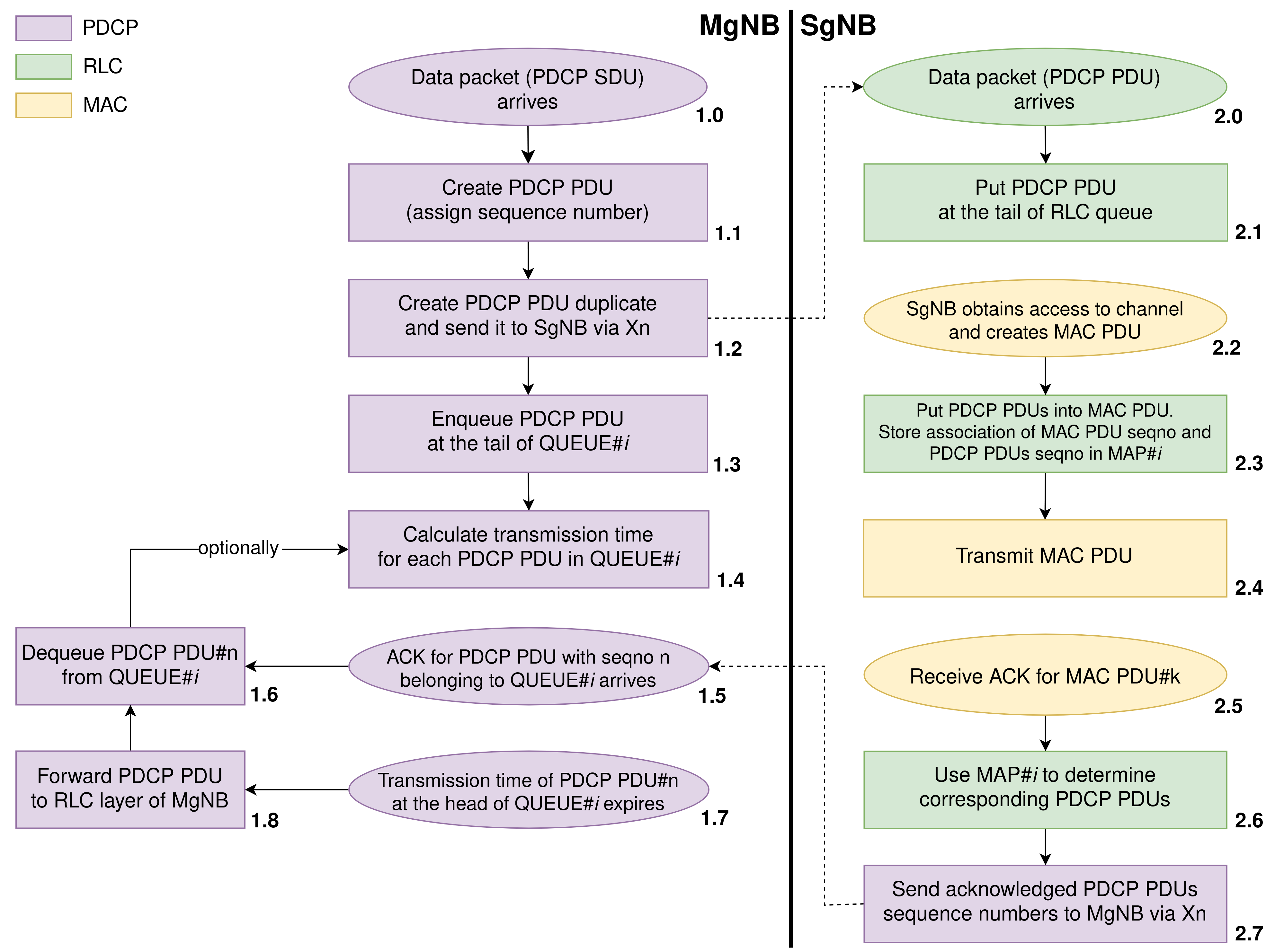} 
	\caption{Flowchart of the proposed \textit{DBTB} algorithm. Different colors denote the layer at which a particular action is performed.}
	\label{fig:buff-scheme}
\end{figure*}

\section{Problem Statement}
\label{sec:problem}
In the paper, we consider the following scenario. An AR/VR application periodically generates video frames. Each video frame is divided into several packets that are sent to the cellular network. The network shall transmit all the packets of a video frame within a given delay budget denoted as $D^{QoS}$. As mentioned in Section~\ref{sec:intro},  $D^{QoS}$ is of the order of a dozen of milliseconds. If at least one packet of a video frame is not delivered within $D^{QoS}$, the whole video frame is considered to be lost. To provide high quality for the AR/VR video at the UE, the fraction of lost video frames shall not exceed $FLR^{QoS}$, which is of the order of $10^{-2}$. Such a low value means that almost all video frames shall be delivered in time.

The traffic of the AR/VR application can be delivered to the UE via two links (i.e., the FR1 link and the FR2 link).
The SgNB operating in FR2 provides high data rates and, thus, can quickly deliver a large portion of data. However, because of specific propagation properties, the data rate can significantly vary with time and  even drop to zero. Thus, a video frame may not be delivered within $D^{QoS}$. In contrast, the MgNB operating in FR1 provides stable but lower data rates.

Our goal is to design an algorithm that delivers video frames within the given delay budget $D^{QoS}$ with the minimal resource consumption of the FR1 link.

\section{Proposed algorithm}
\label{sec:proposed-solution}
This section presents the designed algorithm called \textit{Delay-Based Traffic Balancing (DBTB)}.
The idea behind the DBTB algorithm is as follows. Initially, the algorithm tries to transmit data over the fast FR2 link. Then, if an estimation of the delivery time via the MgNB for some packets exceeds  $D^{QoS}$, these packets are transmitted over the reliable FR1 link. So, the proposed algorithm always uses the fast FR2 link while the reliable FR1 link is left only for ``emergency'' cases, i.e., when packets cannot be delivered in time via the FR2 link.

Below, we describe the actions performed by the MgNB and SgNB (see flowchart in Fig.~\ref{fig:buff-scheme}). 


\subsection{Actions at the MgNB}

First, we describe the actions performed at the MgNB.  When a data packet (PDCP SDU) arrives at the MgNB, the MgNB executes the following actions: (i) assigns a sequence number and creates a PDCP PDU (see Step 1.1 in Fig.~\ref{fig:buff-scheme}), (ii) sends it to the SgNB via the Xn interface (Step 1.2), and (iii) stores its copy (Step 1.3). For that, the MgNB keeps a separate QUEUE$\#i$ for each data stream $i$. 

For the PDCP PDU with a sequence number $n$, the MgNB calculates its transmission time $T_n$ (i.e., the time when the MgNB should start its transmission). If the MgNB does not receive an acknowledgment for the PDCP PDU before time $T_n$ (Step 1.7), it forwards this PDCP PDU to the RLC layer for further transmission (Step 1.8) and deletes it from QUEUE$\#i$ (Step 1.6).

Transmission times $T_n$ in QUEUE$\#i$ are calculated as follows (Step 1.4). For a newly arrived PDCP PDU  $(n+1)$: 
\begin{equation}
	\label{eq:Tn}
	T_{n+1}=T_{now}+(D^{QoS}-D^{ReTX})-s_{n+1}/C,
\end{equation}
where $T_{now}$ is the current time, $s_{n+1}$ is the size of the PDCP PDU $(n+1)$, $C$ is the estimation of the FR1 link data rate, and $D^{ReTX}$ is the time needed to make one retransmission. The protocol stack at the MgNB is configured to provide transmission reliability of $90$\% (i.e., a packet is lost with $0.1$ probability after a single transmission attempt). Thus, to provide data loss lower than $FLR^{QoS}$, for some PDCP PDUs retransmission will be required, which should be taken into account in the overall delay budget.  
Since the addition of a new PDCP PDU increases the time needed to transmit data from QUEUE$\#i$, the MgNB recalculates the transmission times of other PDCP PDU recursively (from tail to head of the queue):
\begin{equation}
	\label{eq:Tn_new}
	T_{n}^{new}=\min(T_n, T_{n+1} - s_n/C).
\end{equation}

In the case of successful transmission of a PDCP PDU via the SgNB, the SgNB sends an acknowledgment containing the corresponding PDCP sequence number to the MgNB (Step 1.5). This PDCP PDU is removed from QUEUE$\#i$ (Step 1.6). Since the number of PDCP PDUs in QUEUE$\#i$ has changed, the MgNB can optionally recalculate transmission times for the remaining PDCP PDUs (Step 1.4).

\subsection{Actions at the SgNB}

Now let us consider the actions performed at the SgNB. All PDCP PDUs arrived from the MgNB are put in the  RLC queue (Step 2.1). The SgNB periodically gets access to the wireless channel and obtains an opportunity to transmit some amount of data called MAC PDU (Step 2.2). In this case, the SgNB can put several PDCP PDUs into the MAC PDU. SgNB stores correspondence between the MAC PDU sequence number and PDCP PDUs sequence numbers into a special structure MAP$\#i$ (Step 2.3). The created MAC PDU is sent to the physical layer for actual transmission (Step 2.4).

The SgNB receives acknowledgments for successfully delivered MAC PDUs (Step 2.5). Upon this event, the SgNB uses  MAP$\#i$ to determine the  delivered PDCP PDUs sequence numbers (Step 2.6). The SgNB uses the Xn interface to notify the MgNB about delivered PDCP PDUs (Step 2.7).



\section{Performance evaluation}
\label{sec:performance-evaluation}


\subsection{Simulation setup}

To compare the performance of the proposed and the existing traffic balancing algorithms, we use the NS-3 simulator~\cite{ns3} with the mmWave module developed in~\cite{ns3-mmwave}. 

We consider a Dense Urban Macro scenario in which UEs are randomly dropped inside a hexagonal cell.  Two gNBs operating in FR1 and FR2 are located at one of the edges of the hexagon. The size of the cell (i.e., the maximum distance between the gNBs and the UE) is $133$~m, which corresponds to the inter-site distance of $200$ m. The main simulation parameters are listed in Table~\ref{table:simulation_parameters}.


\begin{table}[!t]
	\caption{\label{table:simulation_parameters} Simulation parameters.}
	\vspace{-0.4cm}
	\begin{center}
		\begin{tabular}{ |l | c| }
			\hline
			Parameter & Value \\ \hline
			MgNB carrier frequency & $3.6$ GHz \\
			MgNB bandwidth & $100$ MHz  \\
			SgNB carrier frequency & $28$ GHz \\
			SgNB bandwidth & $1$ GHz  \\
			gNBs TX power & $43$ dBm \\
			gNBs height & $10$ m \\
			UE height & $1.6$ m \\
			AR/VR video stream avg. bitrate & $50$ Mbps \\
			Peak to average frame size & 2 \\
			Video frame rate & $60$ FPS \\
			Simulation time & $100$s \\
			Number of simulation runs & $100$ \\
			\hline
		\end{tabular}
	\end{center}
\vspace{-0.9cm}
\end{table}

A server generates an AR/VR video stream and sends it to a UE via a cellular system. The size of each video frame is a random variable defined by the video codec. The video frame should be delivered within delay budget $D^{QoS}=15$~ms. To satisfy the AR/VR QoS requirements, the fraction of frames not delivered in time shall not exceed  $FLR^{QoS}=10^{-2}$.

We implement the solutions described in Sections~\ref{sec:existing-solutions} and~\ref{sec:proposed-solution} and configure them as follows.

\begin{itemize}
	\item \textit{Link Switching}. Every $10$~ms both the MgNB and the SgNB measure SINR in the corresponding channels. The measurements are collected at the MgNB. The MgNB selects the link with the maximum estimated data rate taking into account SINR and the channel bandwidth.
	\item\textit{Single FR1 gNB}. This algorithm corresponds to Link Switching in which the AR/VR traffic is always forwarded to the FR1 link.
	\item \textit{Single FR2 gNB}. This algorithm corresponds to Link Switching in which the AR/VR traffic is always forwarded to the FR2 link.
	
	\item \textit{Packet Duplication}. The MgNB duplicates data packets and sends them over both links.
	\item \textit{Packet Splitting}. Based on the channel measurements, the MgNB determines the data rate of each link $C_{FR1}$ and $C_{FR2}$. For each packet, the MgNB sends the packet via the FR1 link with probability $p=C_{FR1}/(C_{FR1}+C_{FR2})$. Otherwise, the MgNB sends the packet via the FR2 link.
	
	\item \textit{DBTB (Delay-Based Traffic Balancing)}. The Hybrid Automatic Repeat Request (HARQ) protocol at the MgNB is configured such that the target loss probability for a single transmission is $10^{-1}$ and the time needed to make additional retransmission is $D^{ReTX}=5$~ms.
\end{itemize}

To evaluate the performance of the described solutions, we consider the following indicators. 
\begin{itemize}
	\item \textit{Frame Loss Ratio (FLR)} is the fraction of video frames not delivered within  $D^{QoS}$.
	\item \textit{The ratio of satisfied UEs} is the ratio of AR/VR UEs for which FLR does not exceed $FLR^{QoS}$.
	\item \textit{Resource usage} is the fraction of channel resources used for data transmission. Resource usage is measured separately for the MgNB and the SgNB and shows the utilization of the corresponding links.
\end{itemize}

\subsection{Single AR/VR UE}

\begin{figure}[t]
	\centering
	\includegraphics[width=0.9\linewidth]{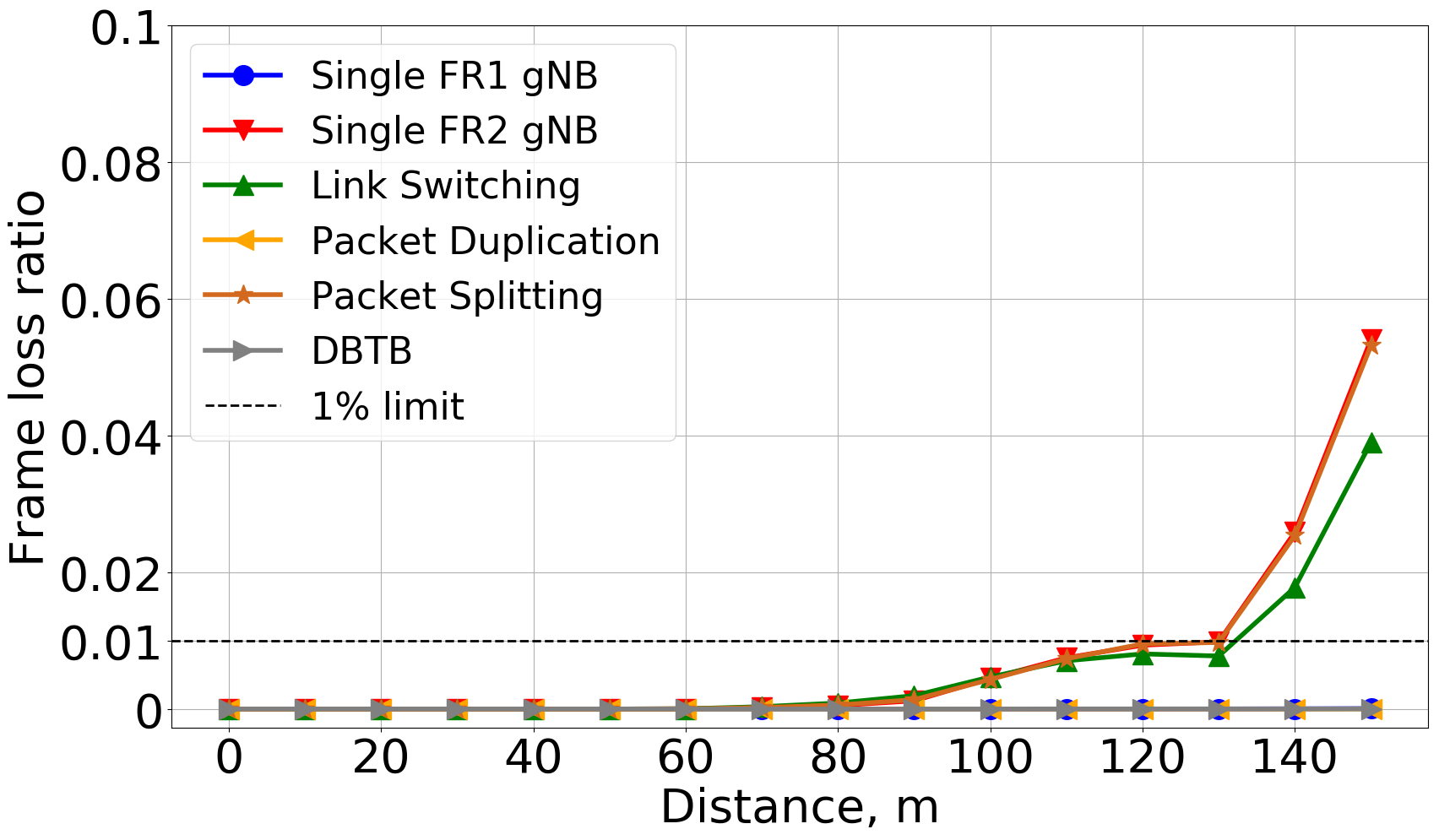} \\
	(a)\\
	\includegraphics[width=0.9\linewidth]{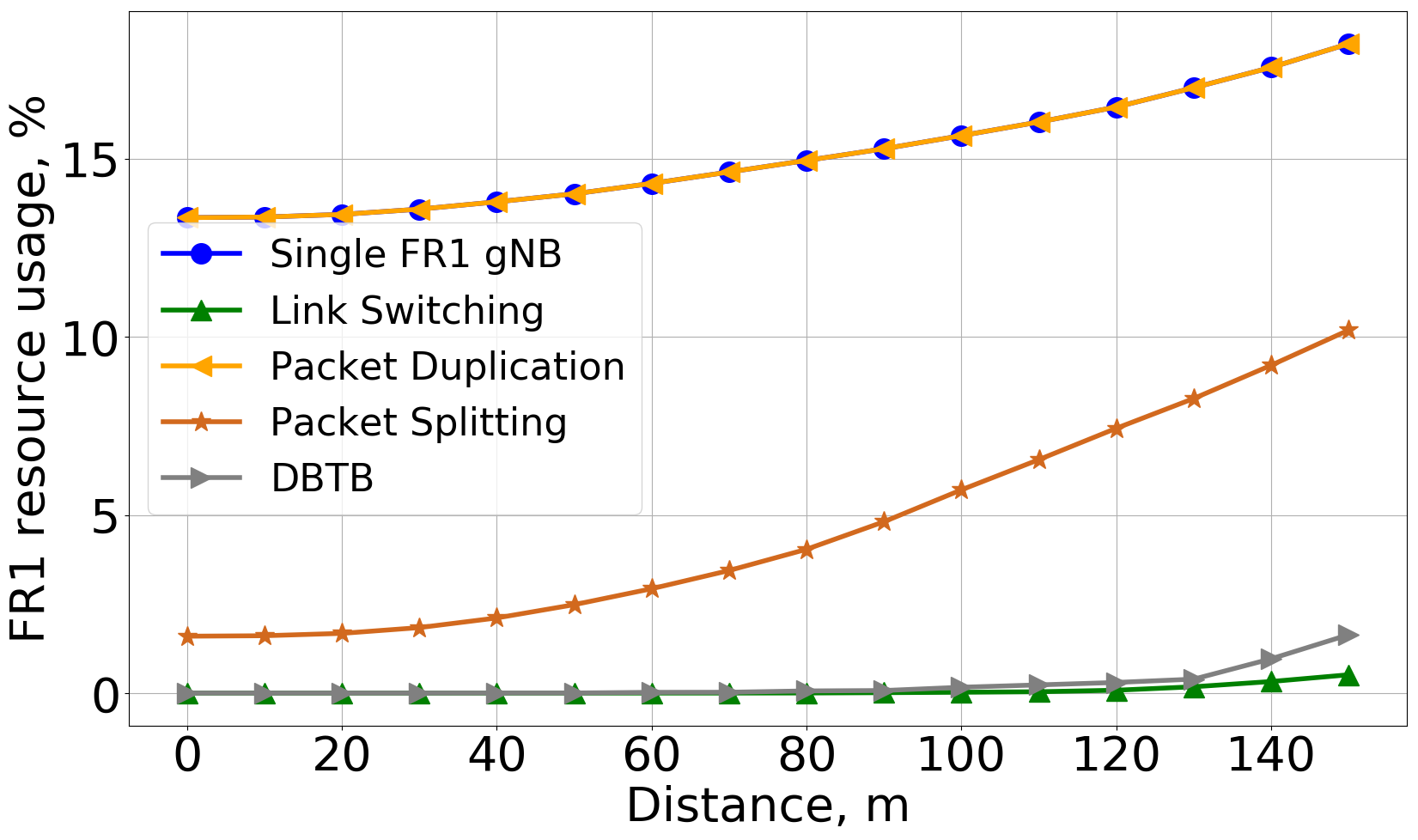} \\
	(b)\\
	\includegraphics[width=0.9\linewidth]{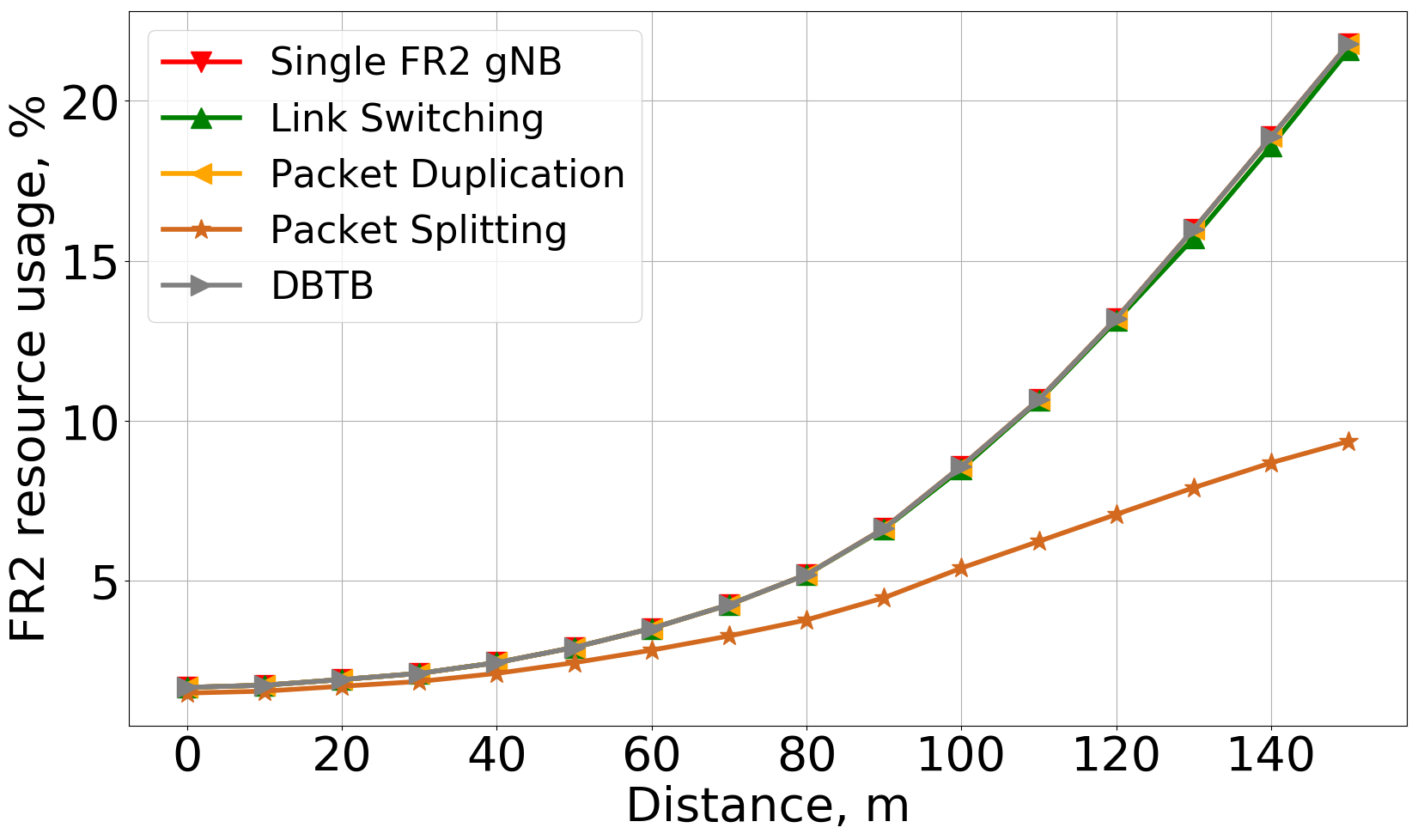} \\ 
	(c)\\
	\caption{Results for the Single AR/VR UE scenario.}
	\label{fig:results-1ue}
	\vspace{-0.4cm}
\end{figure}

In the Single AR/VR UE scenario, we consider a single AR/VR UE located at a fixed distance from gNBs. Fig.~\ref{fig:results-1ue} shows the obtained results. 
We can see that Single FR1 gNB provides FLR below $FLR^{QoS}$ for all distances. 
As shown in Fig.~\ref{fig:results-1ue}(b), a single AR/VR stream occupies approximately 15\% of the FR1 link resources. In contrast,  Single FR2 gNB requires fewer channel resources to transmit the same amount of data because it uses a much wider channel. Since signal attenuation is much higher in FR2 than in FR1, the data rate of the FR2 link significantly decreases with the distance, which significantly increases the channel resource consumption (see Fig.~\ref{fig:results-1ue}(c)). Eventually, Single FR2 gNB does not meet $FLR^{QoS}$ at distances higher than $130$~m. 

The usage of Link Switching and Packet Splitting does not allow increasing reliability in comparison with the Single FR2 gNB because most of the traffic is forwarded via the FR2 link. The reason is that the FR2 link has a much higher average data rate than the FR1 link. Packet Duplication meets the $FLR^{QoS}$ requirement because it always uses the FR1 link. However, it provides the highest resource consumption in both FR1 and FR2 links.

The proposed DBTB algorithm also satisfies the reliability requirement but, in contrast to Packet Duplication, uses a lower amount of the FR1 link channel resources  (less than $2$\%, which is much lower than for Single FR1 gNB). As we will show below, efficient usage of the FR1 link resources is fruitful for other AR/VR streams and other types of traffic that can be served by the MgNB.

\subsection{Multiple AR/VR UEs}

\begin{figure}[t]
	\centering
	\includegraphics[width=0.9\linewidth]{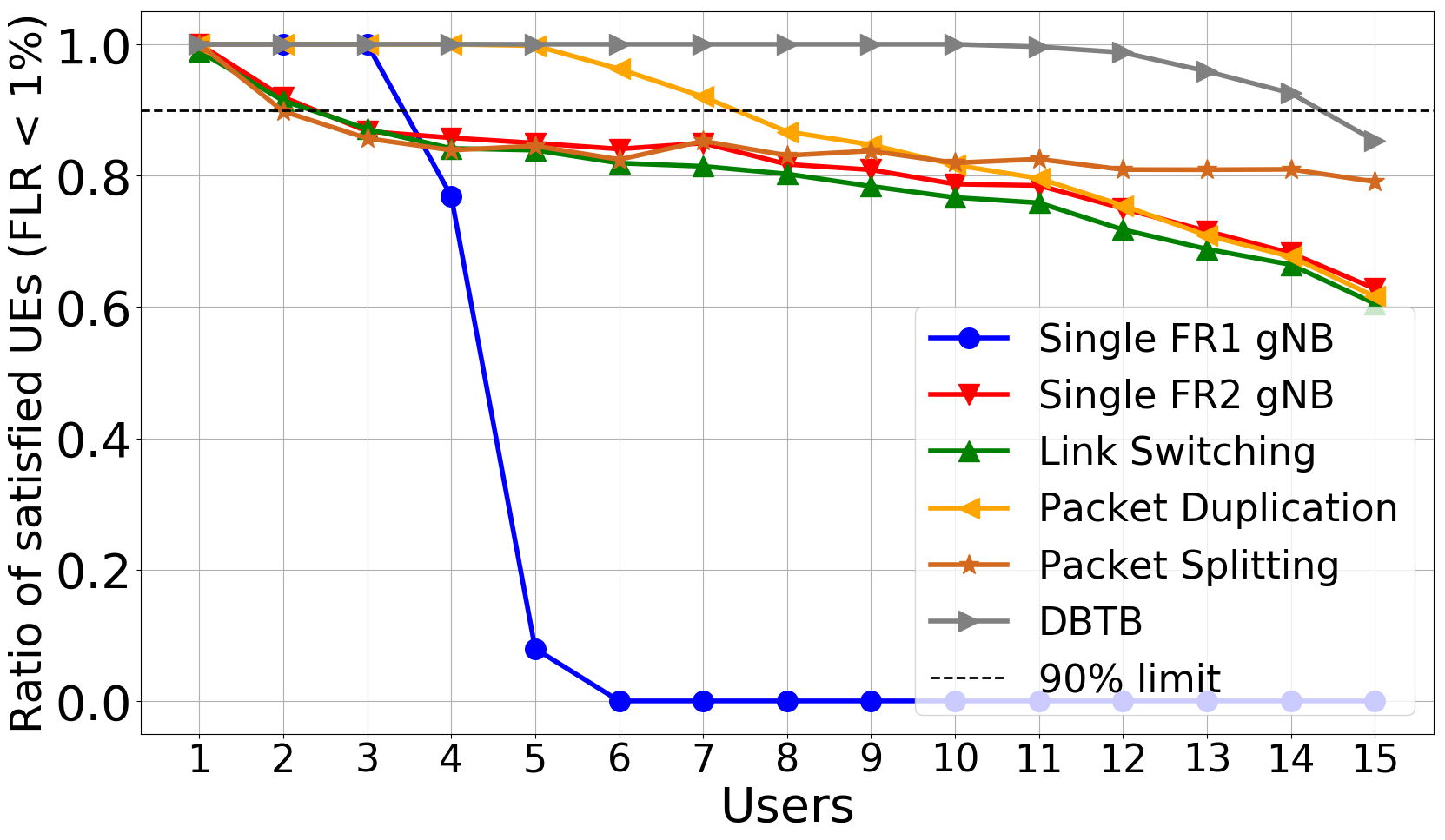} \\
	(a)\\
	\includegraphics[width=0.9\linewidth]{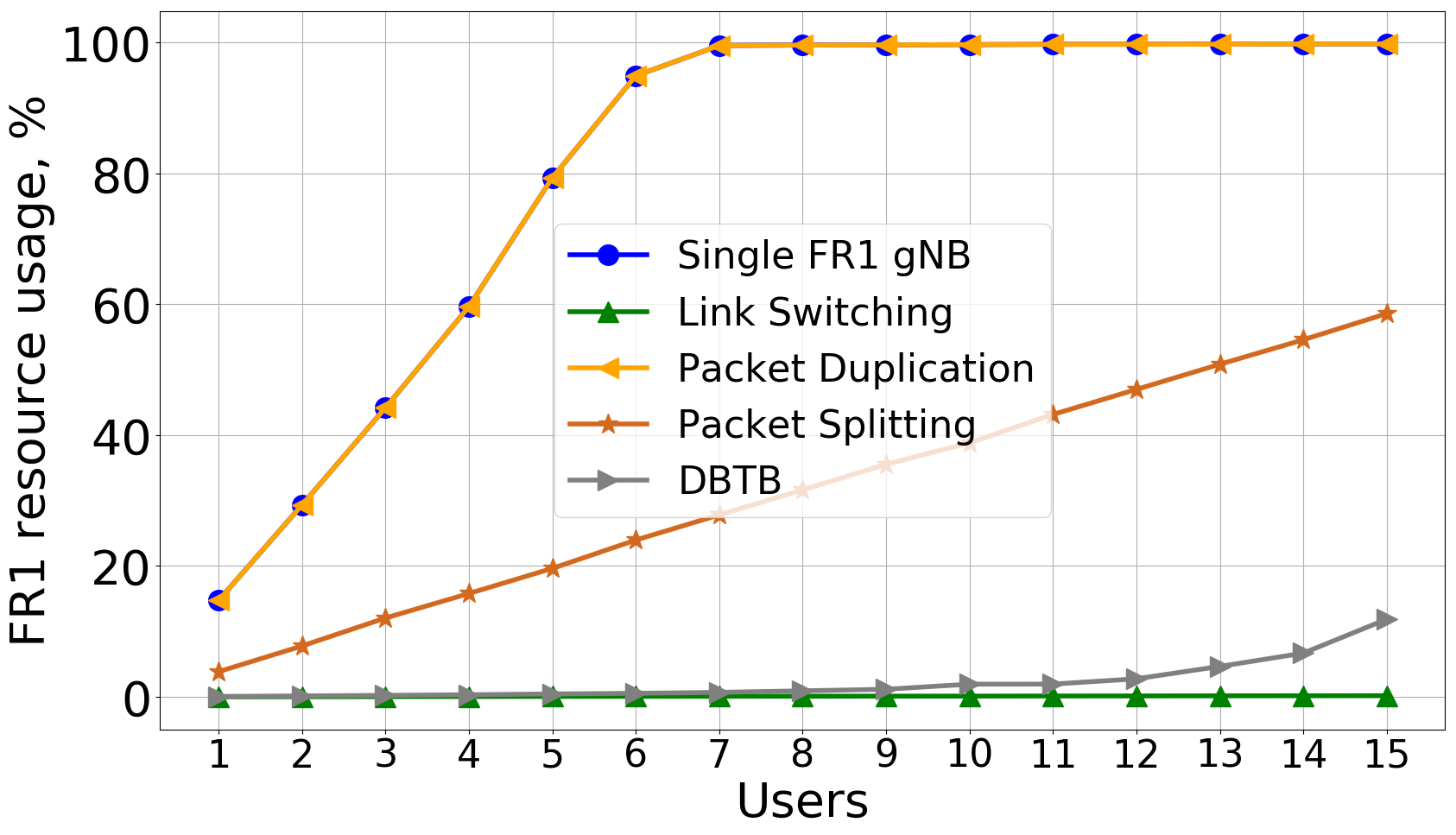} \\
	(b)\\
	\includegraphics[width=0.9\linewidth]{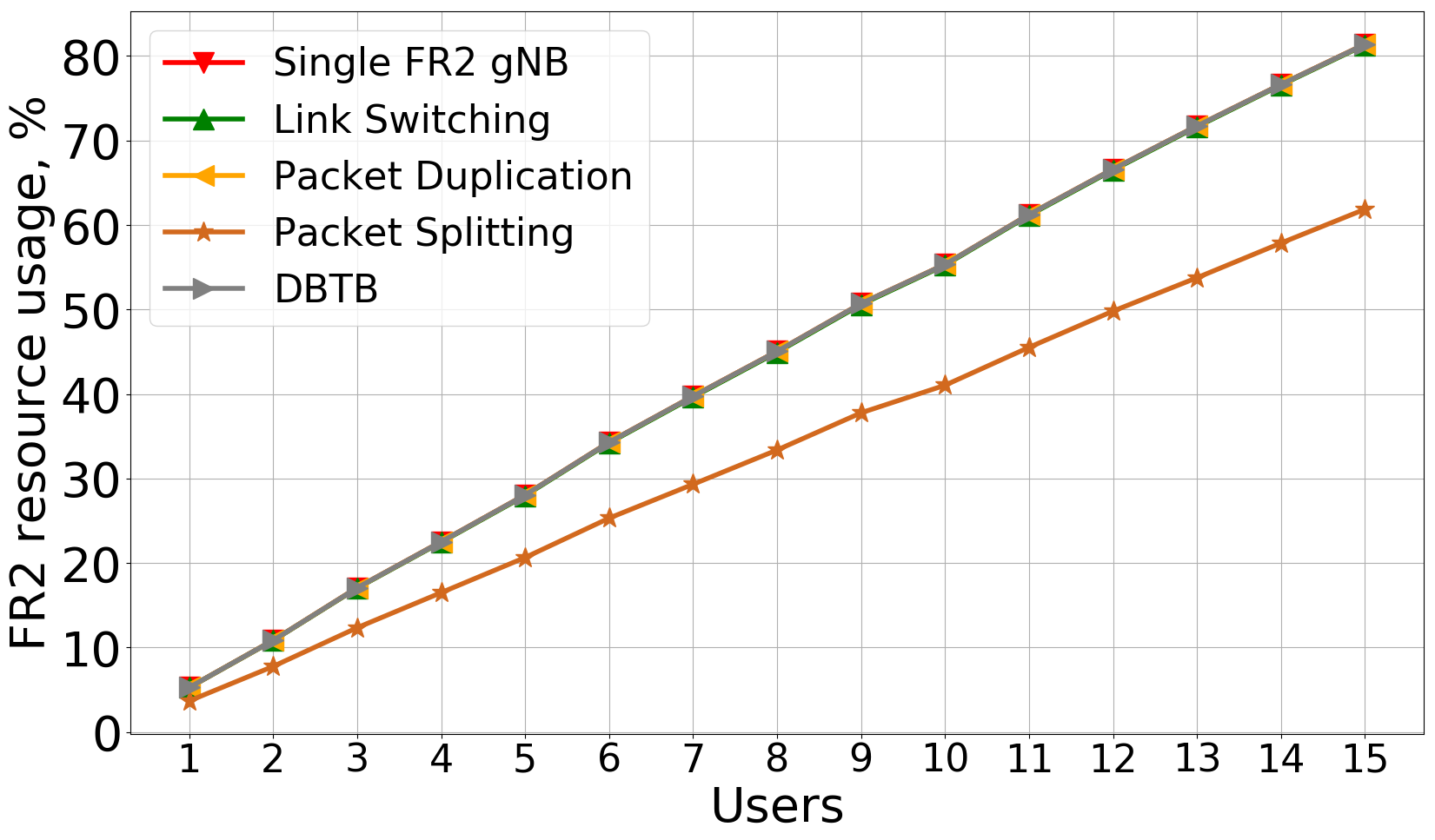} \\ 
	(c)\\
	\caption{Results for the Multiple AR/VR UEs scenario.}
	\label{fig:results-multue}
	\vspace{-0.4cm}
\end{figure}

In the Multiple AR/VR UEs scenario, we consider multiple AR/VR UEs randomly dropped inside a hexagonal cell.
Fig.~\ref{fig:results-multue}(a) shows the ratio of satisfied users as a function of the number of UEs. Following~\cite{tr38.838}, we define the network capacity for AR/VR traffic as the maximum number of UEs for which more than $90$\% UEs are satisfied.
We can see that Single FR2 gNB, Link Switching, and Packet Splitting algorithms provide the network capacity of only 2 UEs. Note that Packet Splitting provides a higher number of satisfied UEs than other listed above solutions because it uses the reliable FR1 link more frequently (see Fig.~\ref{fig:results-multue}(b)).

Single FR1 gNB also provides low capacity (3 UEs) because each AR/VR stream consumes a lot of channel resources and the FR1 link has low bandwidth. Note that the network capacity is reached when the average resource usage is approximately 50\%. The reason is that video frame size is not constant. As detailed in Table~\ref{table:simulation_parameters}, the peak to average frame size is $2$, which means that for transmission of some video frames two times more channel resources are needed. Thus, the operator shall take into account AR/VR traffic burstiness when allocating channel resources. 

The proposed  DBTB algorithm increases network capacity by four times compared to Single FR1 gNB and by two times compared to Packet Duplication. Moreover, we can see that DBTB consumes less than $10$\% of the MgNB resources. Thus, the remaining resources can be used for serving other traffic types. So, we can conclude that the proposed algorithm significantly increases network capacity for AR/VR traffic and efficiently utilizes expensive low-frequency channel resources.



\section{Conclusion}
\label{sec:outro}

In this paper, we considered a scenario in which AR/VR traffic is served in a 5G system that supports Multi-Connectivity (MC). A mobile device is connected to two base stations (gNBs): (i) the first one operates in the low-frequency band (FR1), (ii) the second one operates in the mmWave band (FR2).  
We developed a new algorithm called Delay-Based Traffic Balancing (DBTB) that efficiently uses two links with significantly different characteristics. We compared DBTB with the existing algorithms and showed that DBTB provides significant improvement in terms of the network capacity for AR/VR traffic and resource utilization of the FR1 link.  

In future works, we are going to extend our algorithm for scenarios in which a UE could be connected to multiple gNBs operating in FR2.

\section*{Acknowledgement}
The research has been carried out at IITP RAS and supported by the Russian Science Foundation (Grant No 21-79-10431, https://rscf.ru/en/project/21-79-10431/)


\bibliographystyle{IEEEtran}
\bibliography{biblio}

\end{document}